\documentclass{emulateapj}
\usepackage{times}

\newcommand{\grays}{$\gamma$-rays}
\newcommand{\sigv}{\langle\sigma_a v\rangle}
\newcommand{\pubjournal}[6] {#1 #5, #2, {\bf #3}, #4}

\shorttitle{Dark matter burners}
\shortauthors{Moskalenko \& Wai}

\begin{document}

\title{Dark matter burners}
\author{Igor V. Moskalenko\altaffilmark{1}}
\affil{
   Hansen Experimental Physics Laboratory,
   Stanford University, Stanford, CA 94305 \email{imos@stanford.edu}}
\altaffiltext{1}{Also Kavli Institute for Particle Astrophysics and 
Cosmology, Stanford University, Stanford, CA 94309}
\and
\author{Lawrence L. Wai\altaffilmark{1}}
\affil{Stanford Linear Accelerator Center, Stanford University, 
2575 Sand Hill Rd, Menlo Park, CA 94025 \email{wai@slac.stanford.edu}}

\begin{abstract}

We show that a star orbiting close enough to an adiabatically grown
supermassive black hole (SMBH) can capture weakly interacting massive
particles (WIMPs) at an extremely high rate.  The stellar luminosity
due to annihilation of captured WIMPs in the stellar core may be
comparable to or even exceed the luminosity of the star due to
thermonuclear burning.  The model thus predicts the existence of
unusual stars, essentially WIMP burners, in the vicinity of a SMBH.
We find that the most efficient WIMP burners are stars with degenerate
electron cores, e.g. white dwarfs (WDs); such WDs may have a very high
surface temperature.  If found, such stars would provide evidence for
the existence of particle dark matter and can possibly be used to
establish its density profile.  On the other hand, the lack of such
unusual stars may provide constraints on the WIMP density near the
SMBH, as well as the WIMP-nucleus scattering and pair annihilation
cross-sections.

\end{abstract}

\keywords{black hole physics --- elementary particles ---
radiation mechanisms: non-thermal --- stars: general ---
stars: evolution --- dark matter}

\section{Introduction}

The nature of the non-baryonic dark matter, which dominates the
visible matter by about 4:1, is perhaps the most interesting
experimental challenge for contemporary particle astrophysics.  A hint
for a solution has been found in particle physics where the WIMPs
arise naturally in supersymmetric extensions of the Standard Model
\citep[e.g.,][]{haber-kane}, among other possibilities.  The WIMP is
typically defined as a stable, electrically neutral, massive particle.
Assuming that non-baryonic dark matter is dominated by WIMPs, the pair
annihilation cross-section is related to the observed relic
density \citep{jkg96,bergstrom00}.  A pair of WIMPs can annihilate
producing ordinary particles and \grays.

WIMPs are expected to form high density clumps according to N-body
simulations of test particles with only gravitational interactions
\citep{Navarro97,Moore99}.  The highest density ``free space'' dark
matter regions occur for dark matter particles captured within the
gravitational potential of adiabatically grown SMBHs
\citep{gs99,gp04,bm05}.  Higher dark matter densities are possible for
dark matter particles captured inside of stars or planets.  Any star
close enough to a SMBH can capture a large number of WIMPs during a
short period of time.  Annihilation of captured WIMPs may lead to
considerable energy release in stellar cores thus affecting the
evolution and appearance of such stars.

Such an idea has been first proposed by \citet{salati89} and
further developed by \citet{bouquet89} who applied it to main-sequence
stars. The model led to the conclusion of
suppression of stellar core convection, thus predicting a
concentration of stars in the Galactic Center masquerading as cold red
giants. 

An order-of-magnitude estimate of the WIMP capture rates for
stars of various masses and evolution stages \citep{MW06} lead us
to the conclusion that WDs, fully burned
stars without their own energy supply, are the most promising
candidates to look for. In this paper we calculate the WIMP
capture by WDs located in a high density dark matter region,
and discuss their observational features. We use current limits 
on WIMP-nucleus interaction and WIMP annihilation cross sections, 
as well as recent estimates of WIMP energy density near an 
adiabatically grown SMBH.

\begin{figure*}[t]
\includegraphics[width=0.98\textwidth]{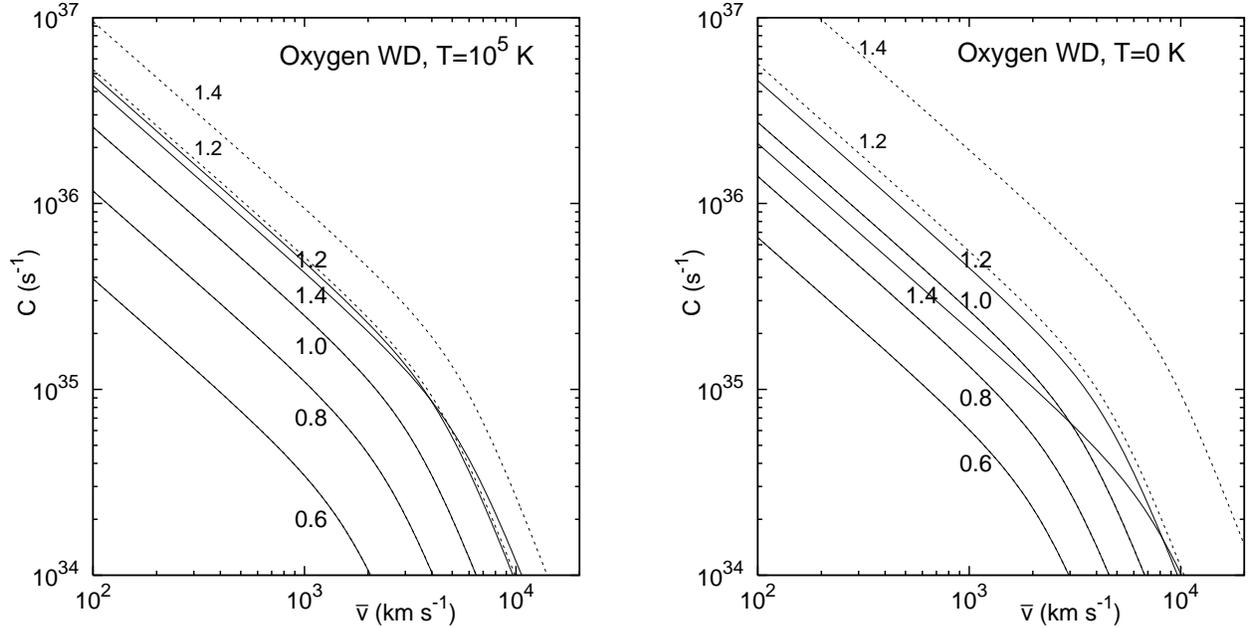}
\caption{WIMP capture rate by Oxygen WDs ($A_n=16$) 
vs.\ velocity dispersion for $\rho_\chi=\rho_\chi^{\max}$.
The left panel shows calculations for WDs
without hydrogen envelope and $T=100,000$ K
\citep{panei00}. Right panel corresponds to the zero-temperature
approximation \citep{hs61}.
The solid lines show the capture rate calculated using the 
modified cross section (Eq.~[\ref{geom_limit}]), 
the dashed lines are calculated for $\sigma'_0=\sigma_0$.
The labels show WD mass in $M_\odot$ units.
\label{C_rate}}
\end{figure*}

\section{WIMP accumulation in stars}

In a steady state the WIMP capture rate $C$ is balanced by the
annihilation rate \citep{gs87}
\begin{equation}
C=A N_\chi^2,
\label{balance}
\end{equation}
where
\begin{equation}
A=\frac{\sigv}{\pi^{3/2} r_\chi^3},
\label{A}
\end{equation}
$\sigv$ is the velocity averaged WIMP pair annihilation
cross-section, the effective radius 
\begin{equation}
r_\chi=c \left(\frac{3T_c}{2\pi G\rho_c m_\chi}\right)^{1/2}
\end{equation}
is determined by matching the star core temperature $T_c$ with the 
gravitational potential energy
(assuming thermal equilibrium), $c$ is the speed of light,
$G$ is the gravitational constant,
$\rho_c$ is the star core density, and $m_\chi$ is the WIMP mass.
The total number of WIMPs captured by a star is
\begin{equation}
N_\chi=C \tau_{eq} \tanh(\tau_*/\tau_{eq}),
\label{Nchi}
\end{equation}
where $\tau_*$ is the star's age, and the equilibrium time scale 
is given by
\begin{equation}
\tau_{eq}=(CA)^{-1/2}.
\label{taueq}
\end{equation}

Limits from direct detection of dark matter on the
WIMP-nucleon cross-section imply that only a fraction of the WIMPs
crossing the star will scatter and be captured.
The capture rate for a Maxwellian WIMP velocity distribution 
(in the observer's frame) 
by a star moving with an arbitrary velocity $v_*$ relative to the
observer is given by \citep{gould87}:
\begin{equation}
C=4\pi \int_0^{R_*} dr\, r^2\, \frac{dC(r)}{dV},
\label{gould2.27}
\end{equation}
where
\begin{eqnarray}
\frac{dC(r)}{dV}&=&
\left(
\frac{6}{\pi}
\right)^{1/2}
\sigma_0 A_n^4 \frac{\rho_*}{M_n}\frac{\rho_\chi}{m_\chi}
\frac{v^2(r)}{\bar{v}^2} \frac{\bar{v}}{2\eta A^2} 
\label{gould2.24}\\
&\times&
\left\{
\left(
A_+A_- -\frac12
\right)
\left[
\chi(-\eta,\eta) -\chi(A_-,A_+)
\right]
\right.\nonumber\\
&+&\left.
\frac12A_+e^{-A_-^2} -\frac12A_- e^{-A_+^2}-\eta e^{-\eta^2}
\right\},\nonumber\\
A^2&=& \frac{3v^2(r)\mu}{2\bar{v}^2\mu_-^2},\nonumber\\
A_\pm&=&A\pm\eta,\nonumber\\
\eta&=&\frac{3v_*^2}{2\bar{v}^2}, \nonumber\\
\chi(a,b)&=&\int_a^b dy\, e^{-y^2}=
\frac{\sqrt\pi}{2}[{\rm erf}(b)-{\rm erf}(a)],
\nonumber
\end{eqnarray}
$\rho_\chi$ is the ambient WIMP energy density,
$A_n$ is the atomic number of the star's nuclei, 
$M_n$ is the nucleus mass, 
$\bar{v}$ is the WIMP velocity dispersion, 
and $\mu=m_\chi/M_n$, $\mu_-=(\mu-1)/2$.
The escape velocity at a given radius $r$ \emph{inside} of a star
is given by
\begin{equation}
v(r)=
\left[
2G\int_{V_*} dV\, \frac{\rho_*(r)}{r}
\right]^{1/2}
=
\left[
\frac{GM_*}{R_*}
\left(
3 -\frac{r^2}{R_*^2}\right)
\right]^{1/2},
\label{vesc}
\end{equation}
where we assumed the same mass density $\rho_*=M_*/V_*$ and the
same chemical composition over the entire scattering volume $V_*$.
This is a reasonable assumption for a degenerate electron core.  Near
a SMBH, where orbital motion around a single mass dominates, the test
particle (WIMP or star) velocities are Keplerian $v_*=\bar{v}$; in
this case $\eta=3/2$, although the exact value does not significantly
change the result. The value of the spin-independent WIMP-nucleon
scattering cross-section $\sigma_0$ is limited by direct detection
experiments, i.e. less than $10^{-43}$ cm$^2$ \citep{cdms06}.  If the
star is composed of nuclei with atomic number $A_n$, the cross
section increases by a coherent factor of $A_n^4$.

If a WD is heavy ($M\ga M_\odot$) and/or $A_n\gg1$, 
almost all WIMPs crossing the star will be captured. 
In this case, the WIMP capture rate is determined
by the geometrical limit $\pi R_*^2$ rather than the total 
interaction cross section $\sigma_0 A_n^4 M_*/M_n$.
We thus use a modified interaction cross section $\sigma'_0$ 
defined as  
\begin{equation}
\sigma'_0 A_n^4\frac{M_*}{M_n}
=\min\left(\sigma_0 A_n^4\frac{M_*}{M_n},\pi R_*^2\right).
\label{geom_limit}
\end{equation}

\begin{figure*}[t]
\includegraphics[width=0.98\textwidth]{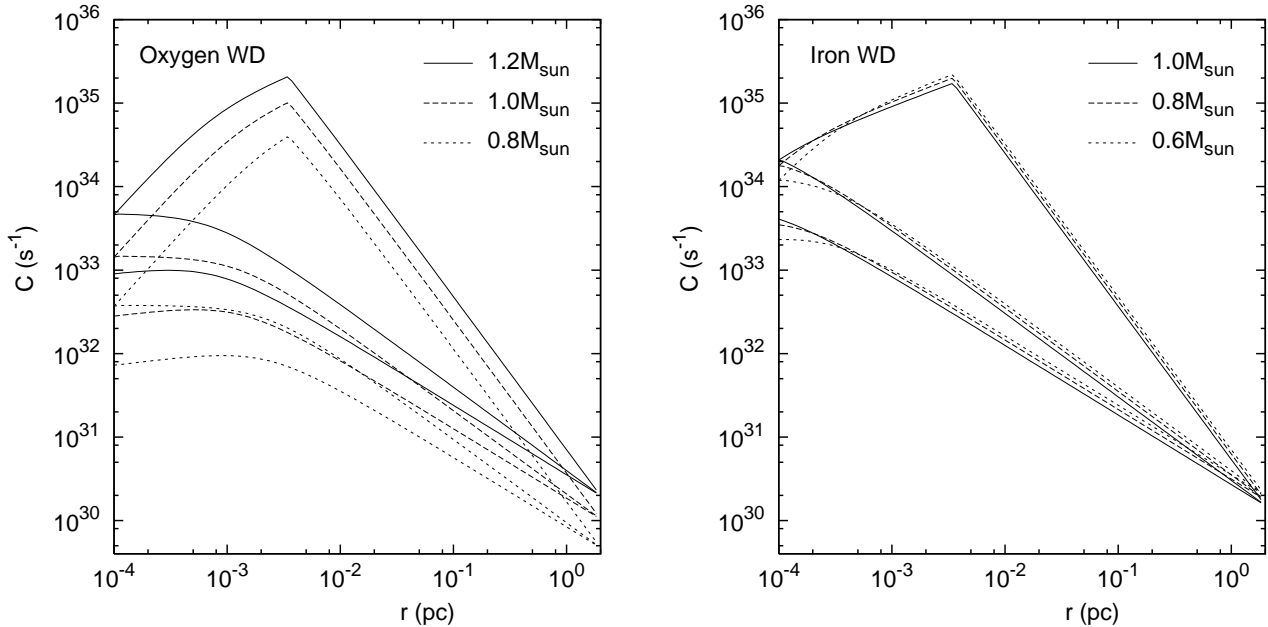}
\caption{Capture rate vs.\ distance from the 
central black hole for Oxygen (left) and Iron (right) WDs of 
$T=100,000$ K assuming Keplerian velocities for 
WIMPs and the WD.
The WIMP mass density is normalized as $\rho_\chi(2\ {\rm pc})=100M_\odot$ 
pc$^{-3}$. For the central spike we assume a power-law profile with the
indices (top to bottom): 7/3, 3/2, 4/3, where the maximal 
possible WIMP mass density $\rho_\chi^{\max}$ is given in the text.
The mass-radius relation is taken from \citet{panei00}.
\label{C_profile}}
\end{figure*}

Figure~\ref{C_rate} shows the capture rate by Oxygen WDs ($A_n=16$)
vs.\ WIMP velocity dispersion for several masses of WDs, assuming
Keplerian orbits around the SMBH, $m_\chi=100$ GeV,
$\sigma_0=10^{-43}$ cm$^2$, $\rho_\chi \sim \rho_\chi^{\max}\sim
m_\chi/(\sigv\tau_{\rm bh}) \sim 10^{10}$ GeV cm$^{-3}$ which
corresponds to the maximal central particle dark matter density
allowed by the age of the SMBH $\tau_{\rm bh}\sim10$ Gyr and our
selected value $\sigv= 3\times10^{-26}$ cm$^3$ s$^{-1}$
\citep{gs99,bm05}. The left panel corresponds to WD effective 
temperature of
$T=100,000$~K (without an envelope), where the masses and radii used
in the calculation are \citep[Fig.~5 in][]{panei00}: $M_*/M_\odot$ =
0.6, 0.8, 1.0, 1.2, 1.4, and $R_*/R_\odot$ = 0.02, 0.012, 0.0085,
0.006, 0.0045, correspondingly.  The right panel corresponds to the
zero-temperature approximation \citep{hs61}, where the mass-radius
relation has been obtained by fitting the numerical results for Carbon
WDs with a function
$R_*/R_\odot=0.94-0.67\tan(1.49[M_*/M_\odot-0.85])$ in the interval
$M_*/M_\odot=0.15-1.4$.  The solid lines show the capture rate
calculated using the modified interaction cross section
(eq.~[\ref{geom_limit}]), and the dashed lines are calculated for
$\sigma'_0=\sigma_0$. In the case of Oxygen WDs the geometrical limit
is reached for $M_*\sim1.2M_\odot$; the larger mass WDs have smaller
radii and therefore smaller capture rates.  For $\bar{v}\la 10^3$ km
s$^{-1}$ our geometrical limit calculations agree well with the
results of \citet{bottino02}, their equation (26).
For $\bar{v}\ga 10^3$ km s$^{-1}$ the
\citet{bottino02} formula, derived under the assumption that each
WIMP crossing the star is captured, gives a systematically larger 
capture rate, up to a factor of 10 for $10^4$ km s$^{-1}$.  
This can be treated as an upper limit, whereas our approximation
to the geometrically limited case can be considered as a lower limit.

It can be seen (Figure~\ref{C_rate}) that cooler WDs have a capture
rate (eqs.~[\ref{gould2.27}],[\ref{gould2.24}]) larger than 
hot ones of the same mass
because of the larger escape velocity (eq.~[\ref{vesc}]). The latter
is the result of a smaller radius $R_*$ and consequently stronger
gravity.  This effect may be explained in terms of the ``focusing
factor'' \citep{gould87} or simply because WIMPs can be captured from
the larger volume of the Maxwellian velocity phase space. A larger
capture rate by a cooler WD will lead to accelerated heating until the
WD radius increases due to increased temperature.  A hotter WD will be
less efficient for WIMP capture and cool down.  This mechanism will
thus lead to fast self-regulation of the WD temperature and capture
rate.

The capture rate for a different ($A_n$) composition WD can be
estimated from scaling the Oxygen WD curves by a factor of
$(A_{n}/16)^3$, e.g., the curves for a Carbon WD can be obtained from
scaling the Oxygen WD curves down by a factor of $(3/4)^3$.  WDs with
heavier nuclei (up to iron) may exist \citep{panei00}; in this case,
the capture rate is restricted mostly by the geometrical limit.

Figure~\ref{C_profile} shows the capture rate for Oxygen (left panel)
and Iron (right panel) WDs vs.\ distance from the central black hole
with $M_{\rm bh}=3.7\times10^6 M_\odot$ \citep{ghez05}; this includes
effects of the radial dependence of the WIMP velocity dispersion and
the WD orbital (Keplerian) velocity.  Following \citet{gp04} and
\citet{bm05}, the WIMP mass density is normalized as $\rho_\chi(2\
{\rm pc})=100M_\odot$ pc$^{-3}$. For the central spike we assume a
power-law profile with indices (top to bottom): 7/3, 3/2, 4/3; these
profiles are predicted for different scenarios of the black hole
growth, adiabatic \citep{ullio01}, quasi-equilibrium \citep{gp04}, and
instanteneous \citep{ullio01}, correspondingly. Here we use the same
estimate for $\rho_\chi^{\max}$ as for Figure~\ref{C_rate}.  The
mass-radius relation for Oxygen and Iron WDs of $T=100,000$~K is taken
from \citet{panei00}.

The capture rate scales linearly with the WIMP density, so that the
largest capture rate is reached with the adiabatic profile.  The
capture rate increases toward the SMBH until the maximal WIMP mass
density $\rho_\chi^{\max}$ is reached; then the capture rate decreases
due to increases in the WIMP velocity dispersion and the orbital
velocity of the star.  For the quasi-equilibrium profile,
$\rho_\chi^{\max}$ is reached only at $\sim$$10^{-4}$ pc, while the
instanteneous profile is even flatter.

As can be seen from Figure~\ref{C_profile} and a simple inspection of
the capture rate formulae, in the geometrically limited case
(eq.~[\ref{geom_limit}]) the capture rate becomes essentially
independent of the WIMP-nucleon scattering cross-section and
degenerate core parameters.  Observationally, the brightest WIMP
burners may be the geometrically limited ones. The main uncertainty in
the geometrically limited capture rate is the dark matter density;
thus it may be possible to perform largely ``model independent"
measurements of the dark matter density profile by measuring the
luminosity of different WIMP burners orbiting within a particular dark
matter spike.

A smaller annihilation cross section $\sigv<3\times10^{-26}$ cm$^3$
s$^{-1}$ would allow for larger ambient WIMP densities near the SMBH.
This would lead to a larger capture rate and consequently larger
burning rate at the innermost radii.  The energy release due to WIMP
annihilation in the stellar core is
$L_\chi\sim 0.16\, C (m_\chi/100\ \rm GeV)$
erg s$^{-1}$ which is actually independent of the WIMP mass
$m_\chi$.

\section{Discussion}

Where does the energy released during the WIMP annihilation go?  Table
1 in \citet{MW06} shows that the effective radius of the thermal
distribution of WIMPs in the stellar core is much smaller than the
radius of the star $r_\chi\ll R_*$, therefore, the products of WIMP
annihilation cannot propagate to the stellar surface and are converted
into thermal energy and neutrino emission.  A WD, a star without its
own energy supply consisting of Carbon and Oxygen, may emit up to
$L_\chi\sim3\times10^{34}$ erg s$^{-1}$, i.e.\ $\sim$10 times the
luminosity of the sun, burning WIMPs only and this energy source will
last forever! (Note that this estimate is based on our \emph{approximation}
to the geometrically limited case and larger luminosities are 
possible even for the given set of parameters.)
At such a luminosity, the surface temperature of the WD
would be close to $\sim$140,000~K, assuming $M_*=1.2M_\odot$, 
$R_*=0.006R_\odot$.  The
maximum of the black body emission falls into the UV band making such
stars strong thermal UV emitters concentrated in the inner $\sim$0.01
pc.  A smaller annihilation cross section $\sigv<3\times10^{-26}$
cm$^3$ s$^{-1}$ and/or larger WIMP density normalization $\rho_\chi(2\
{\rm pc})>100M_\odot$ pc$^{-3}$ would allow for a larger ambient WIMP
density near the SMBH, thus increasing the capture and burning rates
further.

The energy transport in the interiors of WDs is dominated by
degenerate electrons and is very efficient \citep[see][for a recent
review]{hansen04}; therefore, the large number of captured WIMPs and
their annihilation in the core would not change the internal structure
of WDs.  A recently published catalog of spectroscopically confirmed
WDs from the Sloan Digital Sky Survey (SDSS) \citep{e06} contains
several hot WDs with surface temperature in the range of 100,000~K,
thus providing observational evidence that high temperature does not
change the appearance of WDs.  A bare WD with an effective temperature
as high as 170,000--200,000~K has also been observed
\citep[H1504+65,][]{ww99}.

The number of very hot WDs in the SDSS catalog is small, just a
handful out of 9316. This means that observation of a concentration of
very hot WDs at the Galactic Center would be extremely unlikely unless
they are ``dark matter burners.''  The spectra of confirmed hot WDs
can serve as templates for spectroscopic analysis of WDs at the
Galactic Center where only a limited part of the near-IR band can be
used.  An independent determination of the $M_*/R_*$ ratio is possible 
using the gravitational redshift that has to be equivalent
to a radial velocity of about 50 km
s$^{-1}$ \citep{gt67}.

A bare WD with a highly eccentric orbit around the central black hole
may exhibit variations in brightness correlated with the orbital phase
(Figure~\ref{C_profile}).  To have this working, the orbital period
should exceed the equilibrium time scale $\tau_{eq}$
(eq.~[\ref{taueq}]).  Carbon burning stars have $\tau_{eq}\sim10$ yr,
and it is even shorter $\sim$0.5 yr in case of a WD \citep{MW06}.  If
a WD appears in a high-density WIMP region, the WIMP density in its
material would quickly reach equilibrium; thus, the surrounding WIMP
density variation as the WD orbits would result in variations of
brightness.  This makes bare WDs ideal objects to test the WIMP
density in the environment in which they are orbiting.  Since
$L_\chi\propto\rho_\chi$, a population of WDs, bare or with envelopes,
located at different distances from the SMBH would exhibit a
luminosity correlated with the radial WIMP density profile.
Geometrically limited WIMP burners have the highest luminosities and
therefore will be the easiest to observe.  Their luminosity is largely
independent of WIMP-nucleon scattering cross-section, WIMP pair
annihilation cross-section, and degenerate core parameters.

Advances in near-IR instrumentation have made possible observations of
stars in the inner parsec of the Galaxy
\citep{genzel00,ghez03,ghez05}. The apparent K-band brightness of
these stars is 14--17 mag.  The observed absorption line widths imply
high temperatures and lead to a ``paradox of youth:'' apparently young
stars in the region whose current conditions seem to be inhospitable
to star formation.  One of the possibilities is that they are old
stars masquerading as youths. Assuming a central spike with index 7/3,
the K-band brightness for Oxygen WDs with $T\sim100,000$~K and
$R_*/R_\odot\sim0.01$ is about 22--23 mag not including extinction,
which may be as large as 3.3 mag \citep{rrp89}.  It is, therefore,
unlikely that the currently observed stars in the K-band are WDs
burning WIMPs; however, stars with degenerate electron cores plus
envelopes cannot be ruled out.

\acknowledgments

We thank R.~Blandford, J.~Edsj\"o, J.~Faulkner, S.~Kahn, J.~Primack, 
and R.~Romani for interesting discussions and the anonymous referee
for useful comments. L.~L.~W.\ would like
to thank S.~Nagataki for interesting discussions on massive stars.
I.~V.~M.\ acknowledges partial support from NASA Astronomy and
Physics Research and Analysis Program (APRA) grant.  
A part of this work was done at Stanford Linear Accelerator Center, 
Stanford University, and supported by Department of Energy contract
DE-AC03-768SF00515.

\end{document}